# SURFACES REPRESENTATION WITH SHARP FEATURES USING SQRT(3) AND LOOP SUBDIVISION SCHEMES

Yasser M. Abd El-Latif

Department of Math. (Computer Science), Faculty of Science,
Ain Shams University, Abbassia 11566, Cairo, Egypt.

*ABSTRACT*

*This paper presents a hybrid algorithm that combines features form both Sqrt(3) and Loop Subdivision schemes. The algorithm aims at preserving sharp features and trim regions, during the surfaces subdivision, using a set of rules. The implementation is nontrivial due to the computational, topological, and smoothness constraints, which should be satisfied by the underlying surface. The fundamental innovation, in this research work, is the ability to preserve sharp features anywhere on a surface. In addition, the resulting representation remains within the multiresolution subdivision framework. Preserving the original representation has a core advantage that all the applicable operations to the multiresolution subdivision surfaces can subsequently be applied to the edited model. Experimental results, including surfaces coarsening and smoothing, were performed using the proposed algorithm for validation purposes, and the results revealed that the proposed algorithm outperforms the other recent state of the art algorithms.*

*KEYWORDS*

*Computational geometry, Geometric modelling, Sharp features, Subdivision surfaces*

## 1. INTRODUCTION

Subdivision surfaces have been in the highlights of the computer graphics community for the past few years. One of the greatest advantages of subdivision algorithms is that they produce smooth surfaces from an arbitrary control mesh. Nowadays, subdivision algorithms are popular for many kinds of applications, such as mechanical design, geometric modeling, simulation and movie character creation [1]. Many different schemes have been proposed [2], [3], [4], [5], [6], [7]. However, subdivision rules can be derived in the regular case only. Consequently, subdivision surfaces are also famous to exhibit severe artifacts around so called "extraordinary vertices" [8] for which no ideal subdivision rule can be derived. It is interesting to observe that, in all existing schemes, most of the effort concentrated on the design of the smoothing operator. In fact, almost all of them adopt the same standard refinement operators. One characteristic of these refinement operators is that they do not depend on the geometry of the mesh under consideration.

Most of the geometric models include special regions, which do not need to be smoothed. These regions have special properties, which are called sharp features (sharp edges, sharp faces, corners, etc). In the algorithms of Sqrt(3), after several subdivision steps, the surface experiences enough smoothness to represent a fine shape. But, after few subdivision steps, as a consequence, the sharp features are lost and could not be captured within the new mesh.

Therefore, the sharp edges and corners of the original shape are lost through the sampling process and replaced in the result of the subdivision scheme. Hence, the edge flipping, in the Sqrt(3) method, seems to be fundamentally inappropriate for surfaces with sharp features. In this paper, the proposed algorithm allows to preserve the sharp features in the Sqrt(3) subdivision meshes by combining Loop Subdivision, at certain mesh regions, corresponding to these features. Therefore, both the Loop scheme and the 4-8 schemes [9] were adapted naturally to cope with sharp edges while the Catmull-Clark scheme [10] was excluded. The Sqrt(3) scheme is a little more complicated to adapt. This scheme also moves its existing points, but at the same time, it restructures the connections of the edges, so that an edge is only at its same topological position every two steps.

## 2. RELATED WORK

Subdivision surfaces are not limited to completely smooth surfaces. For example, in the Catmull-Clark scheme [10], sharp edges can be generated by keeping the points of the sharp edges in their original place instead of relaxing them by the normal subdivision rules [1]. The surrounding points keep following the standard rules of the scheme. Earlier Hugues Hoppe [11] described similar approaches for Loop's scheme, for which he also added corners and cusp and conical points. These extensions were further analyzed and formalized by the work of Jean Schweitzer [12].

Hugues Hoppe et. al., [13] presented a general method for automatic reconstruction of accurate, concise, piecewise smooth surface models from scattered range data. Novel aspects of the method are its ability to model surfaces of arbitrary topological type and to recover sharp features, such as creases and corners. The method can be used in a variety of applications, such as reverse engineering — the automatic generation of CAD models from physical objects.
Instead of completely sharp edges, also semi-sharp edges are a desired feature, both for modeling artists and for industrial designers. These semi-sharp edges can be generated in a way similar to the sharp ones. Geri's Game [1] shows an example of sharp and semi-sharp features in Pixar's short animation. Sharp edges are most easily implemented on schemes where the original points and edges do not get replaced by multiple new points, but instead are only moved to relax the scheme. Fully interpolating schemes are also difficult to adapt to allow sharp edges, as the existing points are already kept in their place and the surrounding new points would have to fulfill too many constraints.

In a different approach, when the sharp features are identified on the mesh file, modified subdivision rules may be used to subdivide the mesh in order to obtain sharp features in the limit surface [14]. This is particularly useful for multiresolution editing purposes, where, in order to put a curved sharp edge on the limit surface, the user can simply draw a piecewise linear curve on the base domain. Then, this curve will be subdivided through the modified rules that guarantee its eventual sharpness. Ivrissimtzis [15] has done some elegant work on the support of recursive subdivision, which explicitly shows that Sqrt(3) has fractal support.
Attene M. et. al, [16] introduced a simple and efficient edge-sharpening procedure designed to recover the sharp features that are lost by reverse engineering or by remeshing processes, which use a non-adaptive sampling of the original surface. The procedure starts by identifying smooth edges. Then, it performs six trivial filters that identify chamfer edges, which in turn define the chamfer and corner triangles. The chamfer edges and triangles are subdivided by inserting new vertices and moving them to strategic locations where the sharp feature is estimated through extrapolation of abutting smooth portions of the surface.

## 3. OUR PROPOSED SUBDIVISION ALGORITHM

Because standard subdivision approaches would round off the sharp features, we have developed a new subdivision scheme that preserves the sharpness of sharp faces, sharp edges and corners. Our scheme is based on combining both the Sqrt(3) scheme and Loop scheme, with special rules applied on the boundary of the feature regions.

### 3.1. Definition of the scheme

In order to preserve the integrity of the triangle mesh, we group the triangle faces into two groups. The first is called the smooth-face; the normal face that need to smooth during the subdivision process. The second is called the sharp-face; all three boundary edges are tagged as sharp edges and need to preserve during the subdivision process. Indeed, we will differentiate between the face that has three boundary edges as the sharp-edge and the face tagged as the sharp-face. The sharpness features of the first type occur at the boundary of the face, and it is not need to smooth at the inner of the face. But, in the second type, the sharp feature occurs on both the inner and on the boundary of this face. The two cases are illustrated in Section 5 as experimental results.

To define a subdivision scheme for a mesh, we need to specify rules for computing positions of the new vertices that we insert during subdivision process, according to the face and edge type and rules to update the positions of the existing vertices. The rules that we propose are given in Figure 1 and described as follows:

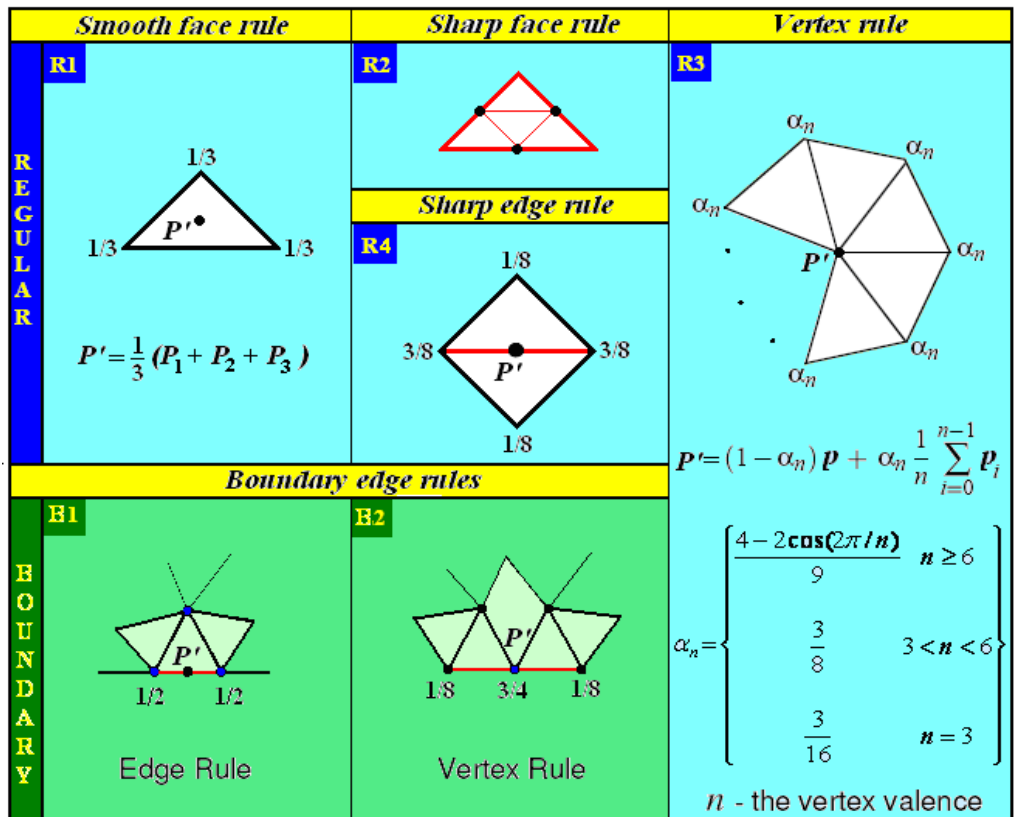

Figure 1. Choose locations for new vertices $P'$ as weighted average of original vertices in local neighborhood for our scheme. Sharp edges are shown in red.

International Journal of Computer Graphics & Animation (IJCGA) Vol.4, No.1, January 2014

**Smooth-Face rule (R1)**: As in the Sqrt(3) scheme, each new vertex is inserted as a result of centeroid refinement of a basic block, i.e., the average point of the vertices of that face;

**Sharp-Face rule (R2)**: As in the Loop scheme, each face split into four new subfaces by inserting three new vertices on each edge, according to the rule (R4) and each vertex updated with the new position;

**Sharp-Edge rule (R4)**: As in the Loop scheme, inserting new vertex on the sharp edge that sharing with at least one sharp face or separate two smooth-faces. For boundaries and edges tagged as *crease* edges, special rules are used (B1, B2). These rules produce a cubic spline curve along the boundary/crease. The curve only depends on control points on the boundary/crease.

**Vertex rule (R3)**: the new position of an existing vertex $P'$ is computed as the average of the old position of $P$, according to the following equation

$$P' = (1-\alpha_n)P + \alpha_n \frac{1}{n}\sum_{i=0}^{n-1} P_i \qquad (1)$$

where $\alpha_n$ is a scaling factor. For $n = 3$ (boundary vertex), $\alpha_n$ is three sixteenths [16]. For $3 < n < 6$ (vertex created on sharp-edges and extraordinary vertex), $\alpha_n$ is three eighths. For $n \geq 6$ (regular and extraordinary vertex), $\alpha_n$ is as Sqrt(3) subdivision scheme.

The basic idea of our proposed algorithm is as follows: if the type of the face is smooth-face, then apply the smooth-face rule, according to R1 rule. Otherwise, apply sharp-face rule, according to R2 rule. If the edge is sharp-edge, then apply R4 rule. The modification of the algorithm is to insert new vertices on each edge tagged as sharp and inside the corner triangles also where several sharp features meet. After few subdivision steps, the generated faces are tagged as smooth and non-smooth faces. Those non-smooth faces preserve the sharp features, during the subdivision scheme. The new algorithm optimizes the surface continuity without losing the sharp features in the special regions, during the construction process, and increases the smoothness of the generated surface in the other regions.

### 3.2. Algorithm Steps

Our algorithm assumes that all of the sharp edges have been identified and tagged. We wish to smooth the triangle mesh to bring it closer to the original curved surface. Because we assume that the vertices are not lie on the original surface, we use an approximation subdivision scheme.

The algorithm can be described as follows. Let $k$ be the number of iterations of the subdivision process and $P^k$ be the polyhedron produced after these $k$ subdivisions. When $k$ is 0, $P^0$ is the initial polyhedron. Let $F$ be the set of all faces that construct $P_i^k$ and S be the set of sharp faces, $S \subseteq F$.

1. For each face $F_i^k (\notin S)$ of $P_i^k$, a new point $V_j^{k+1}$ is made up by the smooth-face rule.

    If $F_i^k (\in S)$, construct four new subfaces by using sharp-rule and mark these new subfaces as the new sharp faces.
2. For every vertex $V_i^k$ of the polyhedron $P^k$, a new vertex $V_i^{k+1}$, termed image, is created by the vertex-rule.

3. For every sharp edges $E_i^k$, a new vertex $V_i^{k+1}$ is created by the sharp-edge rule.

4. For each vertex $V_i^k$, if $V_i^k$ is interior point then a new face $F_i^{k+1}$ is made by connecting the images of $V_i^k$ on the faces meeting at $V_i^k$ in clockwise or counter clockwise direction and take the new vertex created on one of the sharp edges with this rotation. Otherwise, $V_i^k$ is boundary vertex then a new face $F_i^{k+1}$ is made by connecting every adjacent boundary vertex belongs to the same face with the corresponding new point $V_j^{k+1}$ of this face.

The polygons generated through this refinement steps become the input set of polygons for the next iteration step.

## 4. SMOOTHNESS AND NON-SMOOTHNESS ANALYSIS

Since our scheme will be local, we need to analyze only a small number of possible cases of the relationship between the new vertex and the topology of its neighborhood. The two cases of primary importance are the regular sites (all vertices are regular), and the extraordinary sites (adjacent to a non-regular vertex.) After several subdivision steps, at most one vertex in the neighborhood has valence not equal to 6, so it is sufficient to analyze behavior of the scheme only on regular and *n*-regular triangulations, with only one extraordinary vertex of valence *n*.
Hence our subdivision scheme produces a surface that its smoothness depends on the smoothness of Sqrt(3) and Loop subdivision schemes. The smoothness of the limit surface using Sqrt(3) is $C^2$ everywhere except for the extraordinary points where it is $C^1$ [4]. The Loop scheme applied for all the regions that contain the sharp features. Then it produces surfaces that are $C^1$-continuous for valences up to 150, including the boundary case, was proved by Schweitzer [12]. When the two schemes are combined at sharp features (edges or faces) then we will concentrate on analysis of the behavior of the scheme near sharp edges. Figure 2 illustrates that the relation between two adjacent faces with three different cases. Case 1: two smooth-faces are common in sharp-edge. Case 2: one of these faces is sharp. Case 3: two faces are tagged as sharp-face. In all cases, and after several subdivision steps, the created vertices are inserted on the each sharp-edge. This procedure can save the feature attributes of the sharpness either for the edges or for the faces and ensure that no flipping for these type of edges.

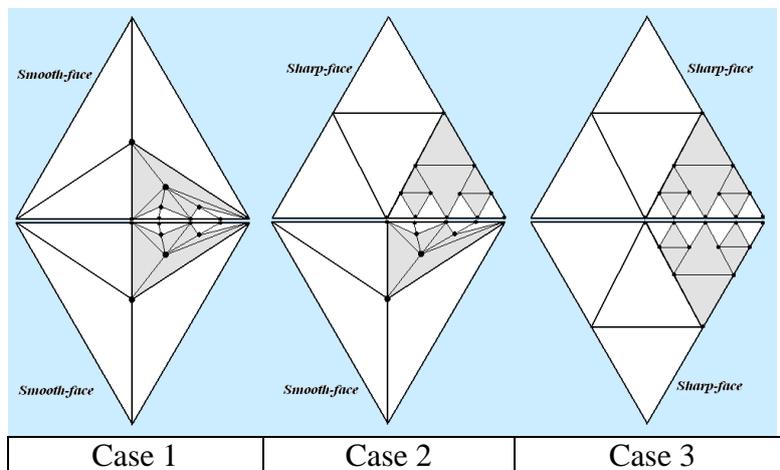

Case 1        Case 2        Case 3

Figure 2. The relation between two adjacent faces with three different cases.

However, itf is often necessary to model surfaces with boundary, which may contain sharp features as well. Thus, it is of practical importance to extend our subdivision scheme to support surfaces with smooth boundaries and creases. Furthermore, it is often useful to have surfaces with piecewise smooth boundary. We have already seen that the subdivision scheme create new vertices of valence 6 in the interior. On the sharp-edges, the newly inserted vertices have valence 4, 5 and 6. On the boundary, the newly inserted vertices have valence 3. Hence, after several subdivision steps, most vertices in a mesh will have one of these valences ($n = 6$ in the interior for smooth-vertices, $3 < n < 6$ in the interior for all vertices that are inserted on sharp-edges, $n = 3$ on the boundary). This means that the surfaces generated using our subdivision are regular except the near of the sharp features and the boundary vertices.

In an extended appendix, we perform a mathematical analysis of the smoothness of the scheme using a combination of regularity and irregularity analysis [12], [13], [17], [18]. Hence the surfaces generated by our scheme are (at least) $C^2$-continuous everywhere [6] except at extraordinary and boundary vertices, where they are $C^1$-continuous. While the surface near curves that bound the sharp face is $C^0$-continuous.

## 5. RESULTS AND COMPARISON

Our subdivision surface representation enables to represent sharp features as internal, boundaries or creases by tagging the control mesh edges. Based on our proposed algorithm, several experimental results with animation have been performed over an arbitrary initial control mesh. The creation of the sharp features and trim regions is illustrated with different levels, from an arbitrary control mesh, using our proposed algorithm. Figure 3 illustrates these rules for different types of faces and edges. It shows how our proposed algorithm applied on sample of mesh surface with sharp features on one triangle (all cases), and its 1-neighborhood. For the smooth mesh, which does not contain any sharp features, we apply Sqrt(3) subdivision scheme for all faces of the mesh. While the mesh surface contains sharp features tagged with red edges, our proposed algorithm will apply Loop subdivision scheme on these edges (faces). We note that the region of sharp edges (faces) and after applied one iteration of Loop subdivision scheme on these edges (faces) the new subfaces does not flipping the edges. Figures 4, 5, 6 and 7 demonstrate the subdivision rule of the proposed algorithm for the surface contains sharp-edges.

Figures 4, 5, 6(a), illustrate the Mannequin, the Teapot and the Cat and its Sqrt(3) subdivision surface algorithms in different subdivision steps. Figures 4, 5, 6(b) show an animation sequence of the same models using our proposed algorithm, respectively. Figure 4(d) illustrates how the sharp-edges will appear with our subdivision algorithm verses Sqrt(3) algorithm by zooming in the ear, and the adjacent region of the eye as the special parts of the Mannequin. Other subdivision process of the Teapot and the HeadCat models using the Sqrt(3) scheme compared with our proposed algorithm is shown in Figures 5, 6 (c) and Figures 5, 6 (d), respectively. Figure 7 illustrates other models in (a) that apply our subdivision scheme to obtain the surface in (b) very close to the limit surfaces in (c).

From these experimental results one can see that, in comparison with Sqrt(3) subdivision algorithm, our proposed subdivision algorithm possesses the following merits. First, it produces good-quality surface approximation with save the sharp features during the construction process. Second, the surface continuity is adaptive. Finally, the proposed technique is still efficient in time complexity and simple to implement, consequently, fast enough for an interactive environment. Such advantages are proved in terms of the analysis given in the previous section.

## 6. CONCLUSIONS AND FUTURE WORK

In this paper we present an algorithm for embedding sharp features in Sqrt(3) subdivision.. Our scheme is based on the combination of the Sqrt(3) scheme and Loop scheme, with special rules applied in the boundary of the feature regions. This algorithm automatically identifies these sharp features and replaces them with refined portions of the mesh that more accurately approximate the original shape. This edge-sharpening process works well for meshes generated by various kinds of uniform samplings and does not introduce undesirable sideffects away from sharp features. We can apply Loop [4] subdivision scheme for the region of mesh that contains the sharp features.

The original contribution and advantages of the proposed algorithm compared with previous techniques are that, it produces good-quality surface approximations without too many faces during the construction process, simple and straightforward to implement, reliable in which it can be applicable to any surface mesh, has certain desirable properties, and its computations are fast enough for an interactive environment.

Our future work will focus on extending our proposed algorithm based on multi-resolution subdivision surfaces [20]. Multi-resolution subdivision surfaces are a natural extension of subdivision surfaces that accommodates editing of details at different scales, allowing general shape deformations as well as the creation of minute features.

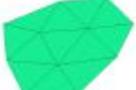

Figure 3. Comparison between applied Sqrt(3) scheme and our subdivision scheme with sharp features on one triangle (all cases) and its 1-neighborhood.

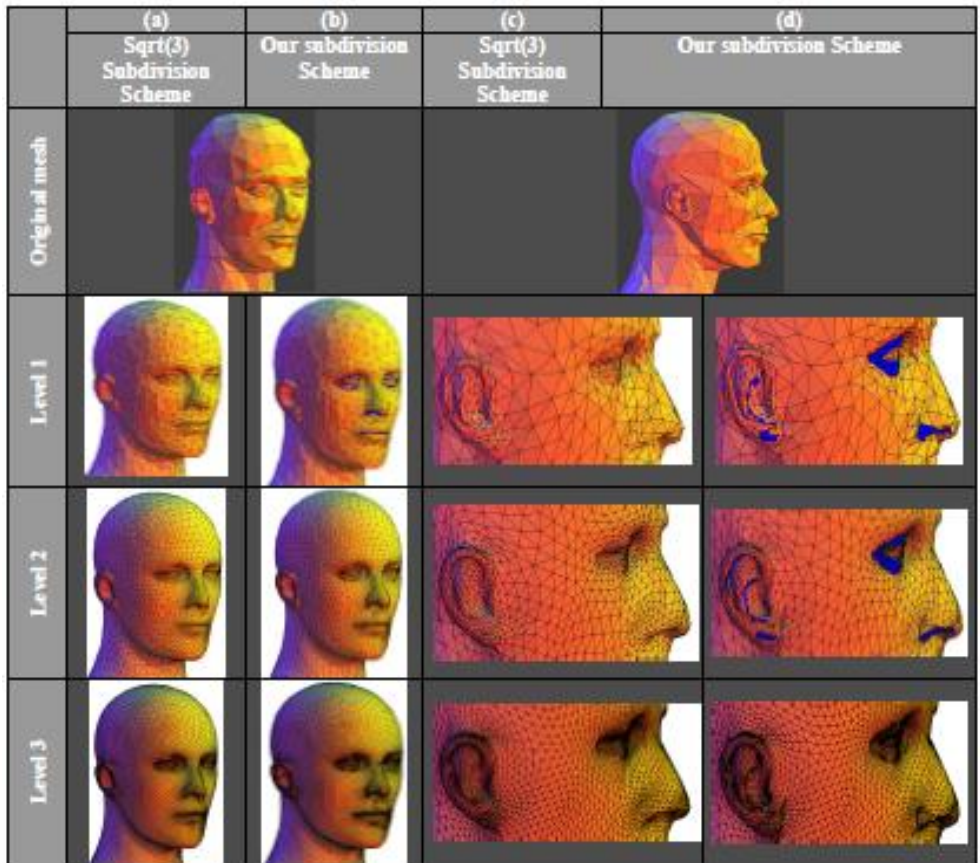

Figure 4. An illustration of the subdivision process of the Mannequin model using
(a) Sqrt(3) and (b) Our proposed subdivision algorithm.

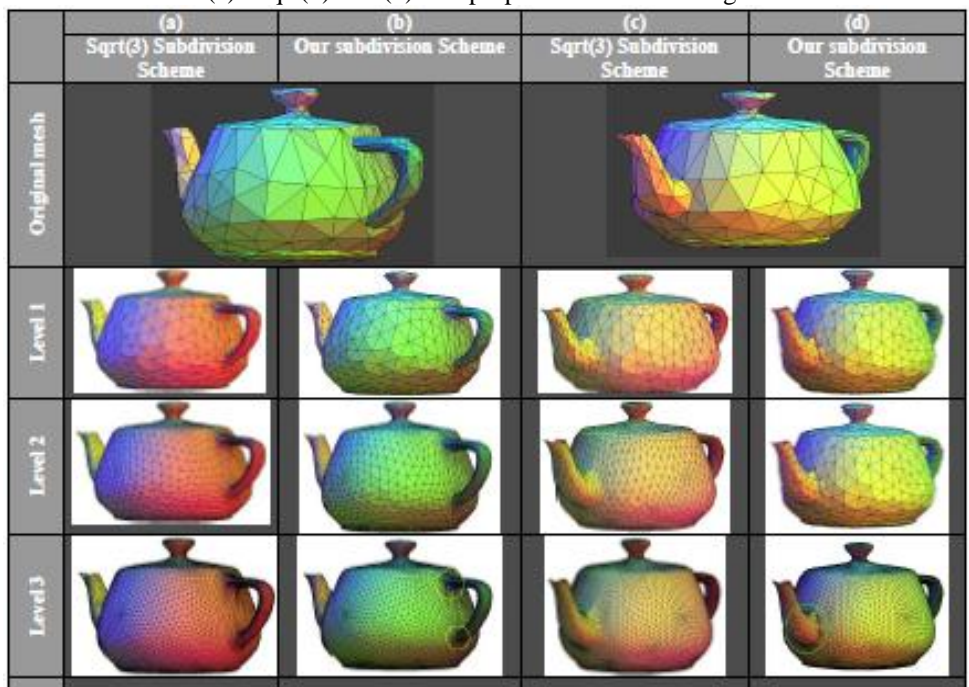

Figure 5. An illustration of the subdivision process of the Teapot model using
(a) Sqrt(3) and (b) Our proposed subdivision algorithm with red sharp-edges.

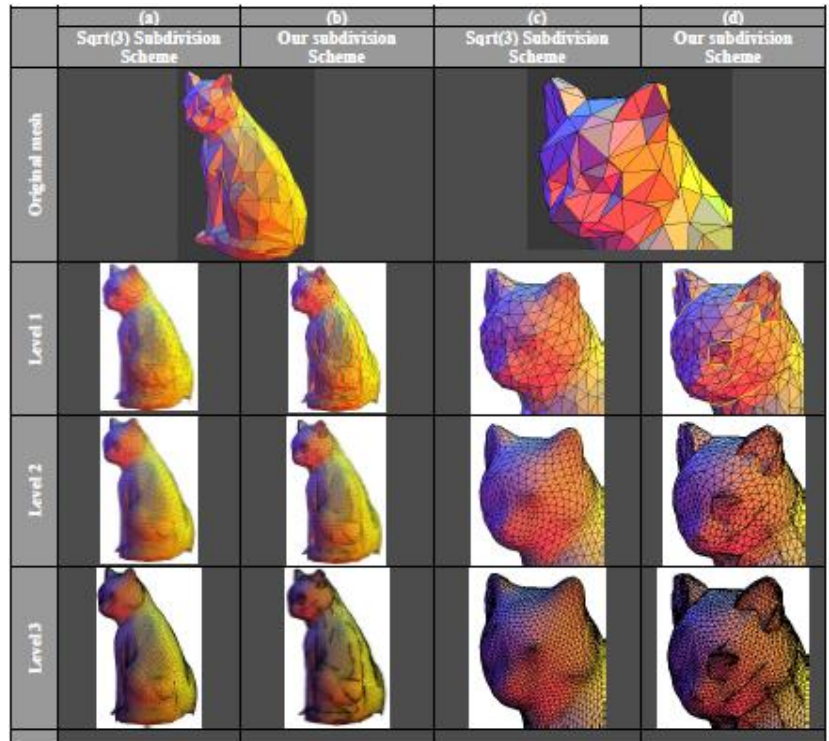

Figure 6. An illustration of the subdivision process of the Cat model using
(a) Sqrt(3) and (b) Our proposed subdivision algorithm with red sharp edges.

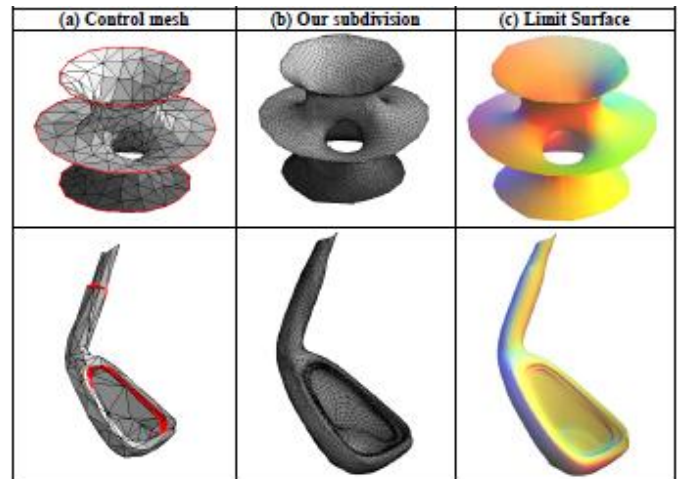

Figure 7. Our subdivision was made to show a closer approximation to the smooth limit surface with sharp features.

## APPENDIX: SMOOTHNESS ANALYSIS

Let $S$ be the subdivision scheme which maps control vertices $p(k)$ from the $k^{th}$ refinement level to the $(k+1)^{st}$ refinement level $p(k+1) = S\ p(k)$. If we consider the action of $S$ on a local neighborhood $V$ only, we can represent $S$ by a matrix with each row containing an affine combination that defines the position of one new control vertex.

We know that applying the Sqrt(3)-subdivision operator two times corresponds to a tri-adic split. So instead of analyzing one single subdivision step, we can combine two successive steps since after the second application of *S*, the neighborhood of $S^2(v_1)$ is again aligned to the original configuration around $v_1$ as illustrated in Figure 8. The 1-ring neighborhood $[v_1, v_2, \ldots, v_n]$ of a vertex $v_1$ is mapped to itself $[w_1, w_2, \ldots, w_n]$ under application of the subdivision scheme. This is reflected by the matrix *S*. If we compute the $m^{th}$ power of the subdivision matrix, we find in the first row a linear combination of $[v_1, v_2, \ldots, v_n]$ which directly yields $S^m(v_1)$.

$$u_1 = (1 - \alpha_n)v_1 + \frac{\alpha_n}{6}\sum_{i=2}^{n} v_i \quad (2)$$

$$u_j = \frac{v_1 + v_j + v_{j+1}}{3}, \quad j = 2, \ldots, n-1 \quad (3)$$

$$w_j = \frac{u_1 + u_{j+1} + u_{j+2}}{3}, \quad j = 1, \ldots, n-1 \quad (4)$$

$$w_n = \frac{u_1 + u_{n-1} + u_2}{3} \quad (5)$$

Hence, we can set-up an $(n \times n)$ matrix which maps *S* and its *n* neighbors to the next refinement level.

$$\begin{bmatrix} w_1 \\ w_2 \\ w_3 \\ \vdots \\ w_n \end{bmatrix} = \frac{5 - 3\alpha_n}{9} v_1 + \frac{1}{3(n-1)} \begin{bmatrix} v_2' \\ v_3' \\ v_4' \\ \vdots \\ v_n' \end{bmatrix} \quad (6)$$

$$\begin{bmatrix} v_2' \\ v_3' \\ v_4' \\ \vdots \\ v_n' \end{bmatrix} = \begin{bmatrix} \frac{3\alpha_n + n - 1}{3} & \frac{3\alpha_n + 2n - 2}{3} & \frac{3\alpha_n + n - 1}{3} & \alpha_n & \cdots & \alpha_n \\ \alpha_n & \frac{3\alpha_n + n - 1}{3} & \frac{3\alpha_n + 2n - 2}{3} & \frac{3\alpha_n + n - 1}{3} & \alpha_n & \cdots \\ \alpha_n & \alpha_n & \frac{3\alpha_n + n - 1}{3} & \frac{3\alpha_n + 2n - 2}{3} & \frac{3\alpha_n + n - 1}{3} & \cdots \\ \vdots & \vdots & \vdots & \ddots & \ddots & \vdots \\ \frac{3\alpha_n + 2n - 2}{3} & \frac{3\alpha_n + n - 1}{3} & \alpha_n & \cdots & \alpha_n & \frac{3\alpha_n + n - 1}{3} \end{bmatrix} \begin{bmatrix} v_2 \\ v_3 \\ v_4 \\ \vdots \\ v_n \end{bmatrix}$$

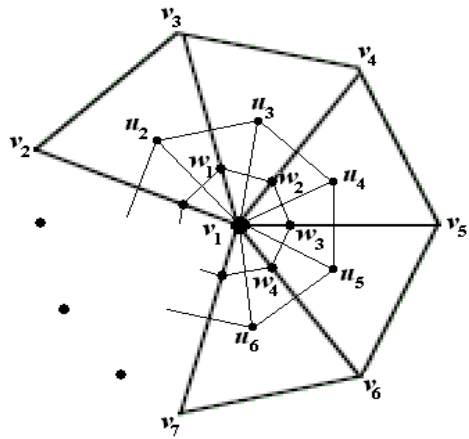

Figure 8. The neighborhood of $S^2(v_1)$ is again aligned to the original configuration around $v_1$.

In particular, our smoothness analysis considers four cases:
- $v_1$ lies on the interior of a smooth-face,
- $v_1$ lies on the interior of a sharp-face,
- $v_1$ lies on the interior of sharp-edge,
- $v_1$ lies at a vertex of the base mesh,

**A. Smooth-faces of the base mesh**

The smoothness of our subdivision scheme on the interior of a smooth-face follows by the smoothness of sqrt(3)-subdivision scheme [6].

Analyzing the action of the Sqrt(3)-subdivision operator on arbitrary triangle meshes, we found that all newly inserted vertices have exactly valence six. The valences of the old vertices are not changed such that after a sufficient number of refinement steps, the mesh has large regions with regular mesh structure which are disturbed only by a small number of isolated extraordinary vertices. Hence the Sqrt(3)-subdivision generates semi-regular meshes since all new vertices have valence six. After an even number $2k$ of refinement steps, each original triangle is replaced by a regular patch with $9k$ triangles. To prove that our scheme is $C^2$ on the given smooth-face of the base mesh, we use the analyzing of the smoothness of triangle subdivision. For Sqrt(3) scheme [9], [25], the resulting matrix has the correct eigenstructure for the analysis. Its eigenvalues are:

$$\frac{1}{9}\left[9, (2-3\alpha_n)^2, 2+2\cos(2\pi\frac{1}{n}), \ldots, 2+2\cos(2\pi\frac{n-1}{n})\right] \quad (7)$$

It is known that for the leading eigenvalues the following necessary conditions have to hold

$$\lambda_1 = 1 > \lambda_2 = \lambda_3 > \lambda_i, \quad i = 4, \ldots, n+1 \quad (8)$$

**B. Sharp-faces of the base mesh**

The smoothness of our subdivision scheme on the interior of a sharp-face follows by the smoothness of Loop-subdivision scheme [4]. In our subdivision scheme, to identify the sharp-faces we use Loop scheme for special region in the mesh. This is because Loop scheme preserves the existing of the edges during the iteration of the subdivision process. While Sqrt(3) scheme flipping the edges with every subdivision process. $C^1$-continuity of the Loop scheme was verified in [12] for valences up to 150.

If we assume regularity of the characteristic map for the subdivision rule, then the subdivision rule will converge to a well defined tangent plane if

$$\lambda_0 = 1 > \lambda_1 \geq \lambda_2 > \lambda_3 \quad (9)$$

For Loops scheme [4], we find the sub-dominant eigenvalues are

$$\lambda_1 = \lambda_2 = (3 + 2\cos(2\pi/n))/8 \quad (10)$$

Since $\lambda_3 = (3+2\cos(4\pi/n))/8$ is smaller than $\lambda_1$ and $\lambda_2$ the condition is satisfied.

## C. Sharp-edges of the base mesh

In the case of the sharp-edge there are three cases as illustrated in Figure 2. Along every sharp-edge there are two types of the vertices. Vertex will inserted as a new one on the middle edge according to the rule (R4), we tagged this vertex of type *B*. Two new vertices will generated as the new position for the old two end vertices for that edge according to the rule (R3), we tagged this vertex of type *A*.

To analyze the smoothness of our scheme along edges of the base mesh is more difficult than the face case since the structure of the vertex whose valence changes between subdivision levels $l$. Let $Val(A_l)$ is the valence of the vertex *A* at level $l$. Figure 9 shows that the valence of the new inserted vertices of type *B* on the interior of sharp-edge and the two end-vertices of type *A* for that edge increased by 2 (for the interior) and by 1 (for the boundary) of the sharp-edge with every level $l$. At level $l \geq 1$, the valence of the vertex of type *A* and *B* is calculated as $Val(A_l) = Val(A_{l-1}) + E_{sh}(A)$ and $Val(B_l) = Val(B_{l-1}) + E_{sh}(B)$, respectively. Where $E_{sh}(V)$ is the number of sharp-edges that incident by the vertex $V$. This means that, the valence of the vertex $V$ on the sharp-edge at level $l$ will increased by the number of incident sharp-edges $E_{sh}(V)$. In practice, we know of no analysis technique capable of establishing the smoothness of our scheme along this edge. However, we can consider that at every subdivision level $l \geq 1$, the vertex of valence not equal to 6 is an extraordinary vertex. Then all we need to analyze the smoothness near the extraordinary vertices.

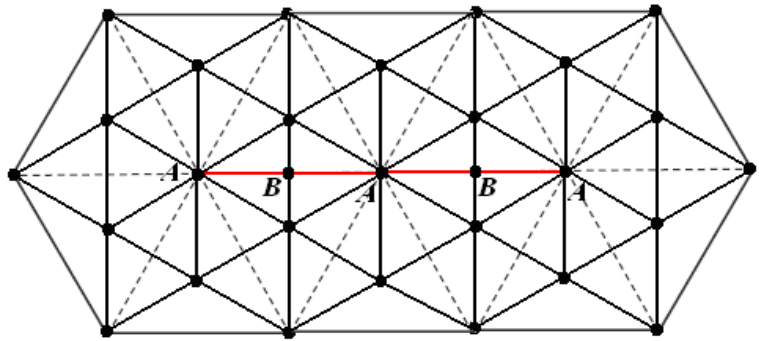

Figure 9. The valence of the new inserted vertices of type *B* = 4 on the interior of sharp-edge and the two end-vertices of type *A* = 7, 8 for that edge.

On regular meshes, subdivision matrices of $C^1$-continuous schemes always have subdominant eigenvalue 1/2. When the eigenvalues of subdivision matrices near extraordinary vertices significantly differ from 1/2, the structure of the mesh becomes uneven: the ratio of the size of triangles on finer and coarser levels adjacent to a given vertex is roughly proportional to the magnitude of the subdominant eigenvalue. To do this a different modification can be used. Rather than modifying the rules for a crease, and making them dependent on the valence of vertices, it is need to modify rules for interior odd vertices adjacent to an extraordinary vertex. For $n \leq 6$, no modification is necessary. For $n > 6$, it is sufficient to use the mask in [19]: instead of 1/2 and 1/4 we can use 1/4+1/4cos(2π/(n-1)) and 1/2-1/4cos(2π/(n-1)) respectively, where *n* is the valence of the extraordinary/boundary vertex. Note that for the Loop scheme the size of the hole in the ring (1-neighborhood removed) is very small relative to the surrounding triangles for valence 3 and becomes larger as *k* grows. For the modified Loop scheme this size remains constant. Then the limit surface can be shown to be $C^1$-continuous at the extraordinary and boundary vertex.

While the continuous of the curves that bound a sharp face is different. Figure 3 suggests these curves can be wiggly and not smooth. This curve is only $C^0$-continuous, and then the surface near such curves also should be.

### D. Vertices of the base mesh

An important aspect of subdivision is the fact that all newly inserted vertices are regular, i.e., their valence is six. Consequently the refined control surface consist of ever larger sections of mesh which are entirely regular with only isolated vertices whose valence is other than six (irregular vertices). In the regular setting when all vertices have valence six the subdivision rules reproduce the refinement rules for quartic box splines. Consequently the limit surface consists of quartic box spline patches almost everywhere. The irregular vertices are at the center of irregular patches which can be thought of as consisting of an infinite geometric sequence of rings of regular patches.

Because of these observations the analytic properties of the limit surface are given by the properties of quartic box splines, except at the irregular vertices. The behavior of the surface at the irregular vertices can be determined by analyzing the local subdivision operator around these vertices, its eigenvalues and (generalized) eigenvectors, and the characteristic map [18]. In the case of the Loop scheme the surface is globally $C^2$ except at the irregular vertices where it is only $C^1$ [12].

The principal result allowing one to analyze $C^1$-continuity of most subdivision schemes is the sufficient condition of Reif [18]. This condition reduces the analysis of stationary subdivision to the analysis of a single map, called the *characteristic map*, for each valence of vertices in the mesh. The analysis of $C^1$-continuity is performed in two steps for each valence:

1. Compute the control net of the characteristic map;
2. Prove that the characteristic map is regular and injective.

The exact condition on the eigenvectors and the injectivity of the corresponding *characteristic map* are quite difficult to check strictly. As mentioned in [6], we therefore restrict ourselves to the numerical verification by sketching the iso-parameter lines of the characteristic map. For completeness we mention that one can develop appropriately modified stencils for boundaries and surface features such as creases and corners [12], [13], [17].

**Yasser M. Abd El-Latif:** He is an Egyptian, birth in Kuwait on 15/04/1971, and graduated from Faculty of Science "pure mathematics and computer science" on 1992. He obtained his MSC in computational geometry in 2000 and PhD in subdivision surfaces in 2005 from Faculty of Science, Ain Shams University, Cairo, Egypt. He is a lecturer of computer science, Faculty of Science, ASU, Egypt. He has been publishing articles in computer graphics and computational geometry. Dr Abd-El-Latif has a membership in Egyptian Mathematical Society (EMS) since 1995, and he is a reviewer in American Mathematical Society (AMS) since 2008 (MR: 57789).
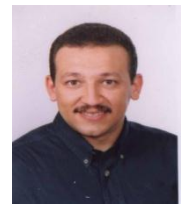